\documentclass[journal]{IEEEtran}
\usepackage{cite}

\ifCLASSINFOpdf
   \usepackage[pdftex]{graphicx}
\else
   \usepackage[dvips]{graphicx}
\fi
\usepackage{amsmath}                            
\usepackage{amssymb}                            
\usepackage{amsthm}

\usepackage{longtable}
\usepackage{array}
\newcolumntype{P}[1]{>{\centering\arraybackslash}p{#1}}
\usepackage{hhline}
\usepackage{algorithm}
\usepackage{algpseudocode}
\usepackage{xparse}
\usepackage{xcolor}

\makeatletter
\NewDocumentCommand{\LeftComment}{s m}{%
  \Statex \IfBooleanF{#1}{\hspace*{\ALG@thistlm}}#2}
\makeatother

\begin{document}
%

\title{PowerGAN: Synthesizing Appliance Power Signatures Using Generative Adversarial Networks} 
%
%
%

\author{Alon~Harell,~\IEEEmembership{Student~Member,~IEEE,}
        Richard~Jones,~\IEEEmembership{Student~Member,~IEEE,}
        Stephen~Makonin,~\IEEEmembership{Senior~Member,~IEEE,}
        and Ivan~V.~Baji\'{c},~\IEEEmembership{Senior~Member,~IEEE}

\thanks{This research was made possible through the NSERC CGS-M scholarship and NSERC Discovery Grants RGPIN-2018-06192 and RGPIN-2016-04590.}
\thanks{Authors are with the \IEEEauthorblockA{Computational Sustainability Lab, School of Engineering Science, Simon Fraser University, Burnaby, BC, Canada} (email:: aharell@sfu.ca, rtj4@sfu.ca, smakonin@sfu.ca, ibajic@ensc.sfu.ca).}
\thanks{Manuscript received June 16, 2020; revised tbd.}}

%
%

\markboth{Submitted for publication pending review}%
{Harell \MakeLowercase{\textit{et al.}}: PowerGAN}
%



\maketitle

\begin{abstract}
Non-intrusive load monitoring (NILM) allows users and energy providers to gain insight into home appliance electricity consumption using only the building's smart meter. Most current techniques for NILM are trained using significant amounts of labeled appliances power data. The collection of such data is challenging, making data a major bottleneck in creating well generalizing NILM solutions. To help mitigate the data limitations, we present the first truly synthetic appliance power signature generator. Our solution, PowerGAN, is based on conditional, progressively growing, 1-D Wasserstein generative adversarial network (GAN). Using PowerGAN, we are able to synthesise truly random and realistic appliance power data signatures. We evaluate the samples generated by PowerGAN in a qualitative way as well as numerically by using traditional GAN evaluation methods such as the Inception score.
\end{abstract}


%
\IEEEpeerreviewmaketitle

\section{Introduction}
\label{sec:intro}
%
%
%
%
\IEEEPARstart{O}{btaining} meaningful insight into the power consumption properties of residential users is a topic of growing importance. Such knowledge allows energy providers to better anticipate future demand, while allowing end users to identify costly appliances within their home, or other energy inefficient habits. Through a better understanding of each specific appliance's power consumption, users and providers  can also begin to reduce the environmental impact of the electric grid. 

Hart~\cite{hart} proposed to determine the power consumption of appliances computationally through what we know as non-intrusive load monitoring (NILM). Using only the smart meter reading of a home, NILM infers the power consumption of appliances within by way of some machine learning or optimization algorithm. 
In most cases, algorithmically, the biggest challenge in NILM is obtaining good approximations of the distributions of appliance power consumption. More specifically, each appliance's posterior distribution conditioned on the aggregate power measurement - $\rho \left(p_i\mid p_H\right)$. 
While unsupervised methods exist for this estimation, such as~\cite{unilm,low_complexity}, it is most  commonly achieved using supervised learning methods. In a supervised setting for NILM, measurements of the aggregate and appliance specific power, taken simultaneously over significant periods of time, are used to build a model of the posterior probabilities. Some of the common models include hidden Markov models~\cite{makonin_sshmm, FHMM}, integer programming~\cite{mixedlinear,aidedlinear}, and more recently, deep neural networks~\cite{kelly, murray_icassp, novel, waveNILM, energan, seq2subseq}.

Using such supervised methods means 
an algorithm's performance greatly depends on how well the training data represents the real distributions. In the context of NILM, this means that the training data needs to represent the true distribution of a household's power consumption characteristics. To ensure a good approximation of the real distributions,  as well as a fair evaluation of performance, long term datasets must be used for training and testing.
As a result, since 2011, data collection has been a main focus of NILM research, and has lead to a creation of a many publicly available datasets such as~\cite{ampds2, REFIT, Tracebase, ECO}
. While these datasets continue to advance the development of NILM solutions, each dataset is unique (in terms of duration, sampling frequency, methodology, etc.) and may only provide a small part of the full distribution of power consumption. 

When considering options for enriching NILM data, and in light of the aforementioned challenges, an alternative approach is to generate synthetic data. 
Our contribution is a novel approach for generating truly random appliance power signatures using generative adversarial networks (GAN). Our synthesizer, named PowerGAN, is capable of generating 
realistic appliance power traces in large quantities, with no hand modeling, allowing for the creation of truly random, new appliances. PowerGAN is unlike previous attempts at generating new power data~\cite{ambal,SmartSIM,SynD,ANTgen}, which are based on simple appliance modeling. PowerGAN is also novel within the existing GAN literature, as it presents an improvement over existing time-series generators based on GANs. 

\vspace{-0.3cm}
\section{Related Work}
\label{sec:prev}

\subsection{Generative Adversarial Networks (GANs)}
\label{subsec:GAN}

Until recently, the main use for deep neural networks (DNN) was solving problems such as classification, regression, or segmentation. While DNNs were highly successful at such tasks, including NILM~\cite{kelly, waveNILM}, they were not able to generate synthetic data. This changed in 2014 with the introduction of generative adversarial networks (GAN)~\cite{GAN}. 
The main novelty in GAN is that instead of one neural network trained to solve an optimization problem, two competing neural networks are trained to find the equilibrium of a game. 

\begin{figure}
  \centering
    \includegraphics[width = 0.9\linewidth]{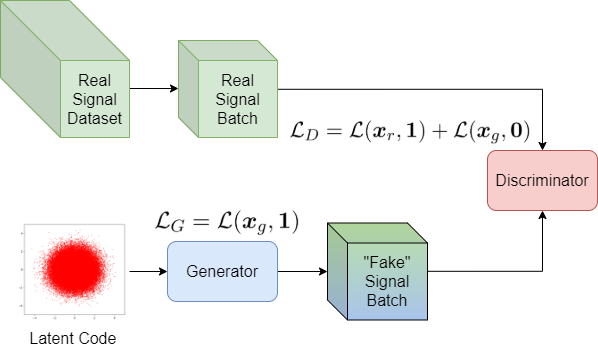}
    \caption{GAN structure with alternating discriminator and generator training.} 
    \label{fig:GAN} 
    \vspace{-0.5cm}
\end{figure}

The two players in the GAN game are known as the \textit{generator} and the \textit{discriminator}. The generator tries to generate realistic signals 
from a random input known as the \textit{latent code}, while the discriminator attempts to successfully distinguish these generated signals from real ones. The training process is performed in turns, alternating between training the generator and the discriminator once (or more) at each turn. Fig.~\ref{fig:GAN} shows a visual explanation of the GAN framework. The equilibrium of the GAN game is achieved when the generator can create perfectly realistic signals, 
so that even a perfect discriminator cannot distinguish them from real ones.

The introduction of GANs allowed DNNs to generate increasingly realistic signals such as faces or scenes~\cite{dcgan}. However, basic GANs, sometimes known as vanilla GANs, 
remain difficult to train. To improve both the final outcome as well as increase the stability of GAN training, many variations on the GAN framework have been published. Goodfellow \textit{et al.}~\cite{improved} suggested label smoothing, historical averaging, and minibatch discrimination. Arjovsky \textit{et al.}~\cite{wgan, wgan-gp} showed that KL divergence between real and fake sample outputs of the discriminator, the commonly used loss function in GAN training, suffered from vanishing gradients, and suggested using the Wasserstein distance instead. The corresponding GANs are referred to as Wasserstein GANs (WGANs). 
Gulrajani \textit{et al.}~\cite{wgan-gp} presented the gradient penalty as a way to increase the stability of WGAN training. 
Other improvements include using a conditional generator based on class labels~\cite{cgan,acgan}, and conditioning the generator on an input signal ~\cite{cyclegan} to transform the output.  

Basic GANs, mentioned above, are limited in performance as well as difficult to train. This makes vanilla GANs insufficient for the challenging task of representing the true distributions of appliance level power signatures. When approaching the development of our own GAN model, we considered two specific 
versions of GAN -- Progressively growing GAN~\cite{ProgGAN}, and EEG-GAN~\cite{eeggan}, both of which use the WGAN loss with gradient penalty as the underlying GAN loss.

Karras \textit{et al.}~\cite{ProgGAN} have shown that it is beneficial to train GANs in stages. At first, 
coarse structure is learnt by training a GAN on highly downsampled signals. After sufficient training, the next stage of the GAN is added and the signal resolution is doubled. At this stage the weights that had previously been learnt are kept and additional layers are added. On the generator side, the layers are added at the end; whereas, on the critic side they are added at the beginning.

In~\cite{eeggan} Hartmann \textit{et al.} present EEG-GAN, an adaption of~\cite{ProgGAN} for the generation of electroencephalogram signals. The training algorithm closely resembles that of~\cite{ProgGAN}, with modified architectures for generating 1-D time-series data instead of images. Despite the similarity in training, the authors do present several modifications in EEG-GAN, the combination of which was novel at the time of publication. One of particular importance to PowerGAN is the weighted, one-sided gradient penalty, which is adopted by PowerGAN and expanded on in Section~\ref{subsec:PowerGAN}. 

\vspace{-0.3cm}
\subsection{Power Data Synthesizers}
\label{subsec:synthesizers}

The challenges presented by the available long-term disaggregation datasets have motivated  several efforts to generate synthetic data for NILM. These efforts, varying in sophistication and scope, focus on generating realistic \textit{aggregate} signals. In contrast, the proposed PowerGAN is focused on \textit{appliance-level traces}. Nonetheless, these power data synthesizers all employ some techniques for simulating appliance-level data before layering it to create the aggregate.

SmartSim~\cite{SmartSIM} was one of the first such power data synthesizers. SmartSim's appliance level simulation is performed by matching each appliance with one of four possible energy models: 
ON-OFF, ON-OFF with growth/decay, stable min-max, and random range models. Reasonable parameterizations for each of these models were extracted by the authors from real instances of the specific appliances in the Smart* dataset~\cite{SMART}. 
The estimation of these values directly from real data, taken from the Smart* dataset, inherently limits SmartSim's ability to capture the variability of real 
appliances. Furthermore, by copying these parameters from real data, SmartSIM provides no new appliance-level traces. 


The Automated Model Builder for Appliance Loads (AMBAL)\cite{ambal} and its recent iteration, ANTgen~\cite{ANTgen}, approach appliance models similarly. They employ the same four general appliance classes with the addition of compound model types. Compound models are combinations of the four basic models, and are generally a better fit to real-world appliances. Model parameters are determined using the ECO~\cite{ECO}
and Tracebase datasets~\cite{Tracebase} where active segments of each appliance are broken up according to possible internal state changes. Rather than deciding \textit{a priori} the model class for a particular appliance, AMBAL/ANTgen selects the model fit that minimizes the mean absolute percentage error. 

SynD~\cite{SynD} is a similar effort that instead categorizes appliances as either autonomous or user-operated. Autonomous appliances include constantly-on loads (such as a router) or appliances that are cyclic in their operation patterns (such as a fridge). User-operated appliances can involve single-pattern operation (such as a kettle) or multi-pattern operation (such as a dishwasher or programmable oven). On the appliance level, power traces for SynD were measured directly by the authors and stored as templates. 

The extraction of appliance models directly from real data restricts the ability of these generators to provide truly novel appliance-level traces. However, the aim of these generators is to synthetically expand the space of realistic aggregate signals, which has and will continue to contribute to the NILM community. In contrast, our work focuses on appliance-level modeling, moving past 
the parameterization of pre-specified appliance models, and instead making use of the rapidly developing generative-adversarial framework to elucidate entire distributions over appliance behaviour. Note that we do not compare with SHED~\cite{SHED}, which uses similar methods, because it is designed for 
commercial buildings rather than residential ones.

It is also important to note that GANs have been used for NILM in~\cite{bao2018enhancing, energan, seq2subseq}. In~\cite{bao2018enhancing} a pretrained GAN generator is used to replace the decoder side of a denoising autoencoder based disaggregator. In~\cite{energan, seq2subseq}, GANs were heavily conditioned on aggregate data and simply used as a refinement method for supervised disaggregation using convolutional neural networks. However, none of these works use GANs for the purpose of generating new data, evaluate their models using conventional GAN metrics, or made their models publicly available, and as such are not comparable with PowerGAN.

\vspace{-0.3cm}
\section{Methodology}
\label{sec:method}

\subsection{PowerGAN}
\label{subsec:PowerGAN}

Both progressive growing of GANs and EEG-GAN introduce novel methods of training GANs, with a variety of techniques for improved performance and reliable convergence. However, neither of the two methods takes advantage of class labels. Inspired by~\cite{acgan,cgan}, we extend EEG-GAN by conditioning both the generator and the critic on the specific appliance label. We name our framework PowerGAN - a conditional, progressively growing, one dimensional WGAN for generating appliance-level power traces. 

\begin{figure}
  \centering
    \includegraphics[width = \linewidth]{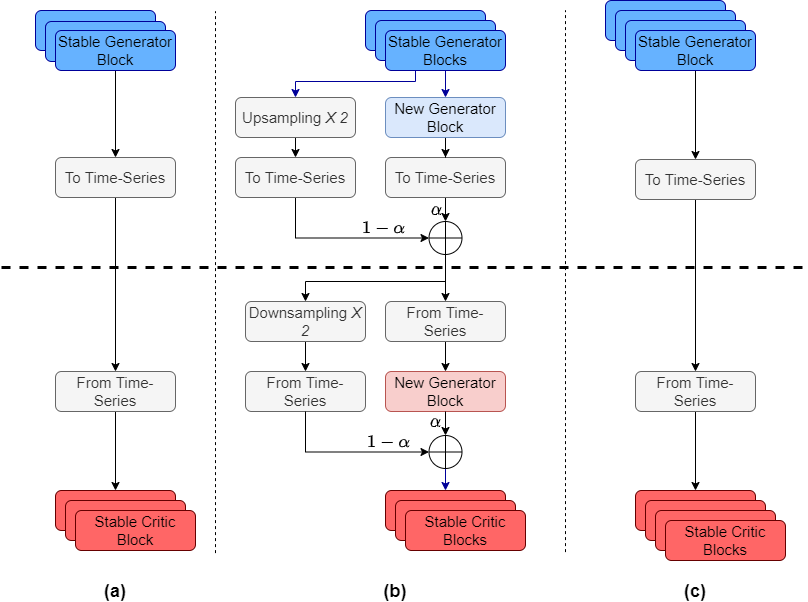}
    \caption{The fading procedure proposed by~\cite{ProgGAN} as adapted for one time-series data in~\cite{eeggan} and PowerGAN. In (a) we see the currently stable generator and critic during an intermediate stage of training; note that 
    generator (critic) 
    contains a upsampling (downsampling) step. 
    The blocks ``To Time-Series" and ``From Time-Series" are implemented via 1D convolution. In (b) we see the fading stage. On the generator side, the output of new blocks is slowly faded in, using a linearly growing parameter $\alpha$, with an nearest neighbor upsampling of the output of the stable blocks. Similarly, on the critic side, the features created by the new block are slowly merged in with previous inputs to the existing critic blocks. Finally, (c) shows the blocks after the fading is complete and $\alpha = 1$. In PowerGAN, this fading is performed over 1000 epochs, allowing for knowledge obtained at earlier steps of training to slowly adapt as new layers are added.} 
    \label{fig:fading} 
    \vspace{-0.5cm}
\end{figure}

The basic architecture of PowerGAN is similar to the EEG-GAN adaptation of \cite{ProgGAN}. PowerGAN contains six generator and critic blocks, each comprised of two convolutional layers and an upsampling, or downsampling layer respectively.
Following the process in~\cite{eeggan,ProgGAN}, we perform a fading procedure each time a new block is added. During fading, the output of a new block of layers is scaled by a linearly growing parameter $\alpha$ and added to the output of existing layers which is scaled by $1-\alpha$. All layers remain trainable throughout the process and the corresponding dimensionality discrepancies are resolved by a simple $1\times 1$ convolutional layer. An illustration 
of this 
process 
is shown in Fig.~\ref{fig:fading}.

\begin{figure*}
  \centering
    \includegraphics[width = 0.9\textwidth]{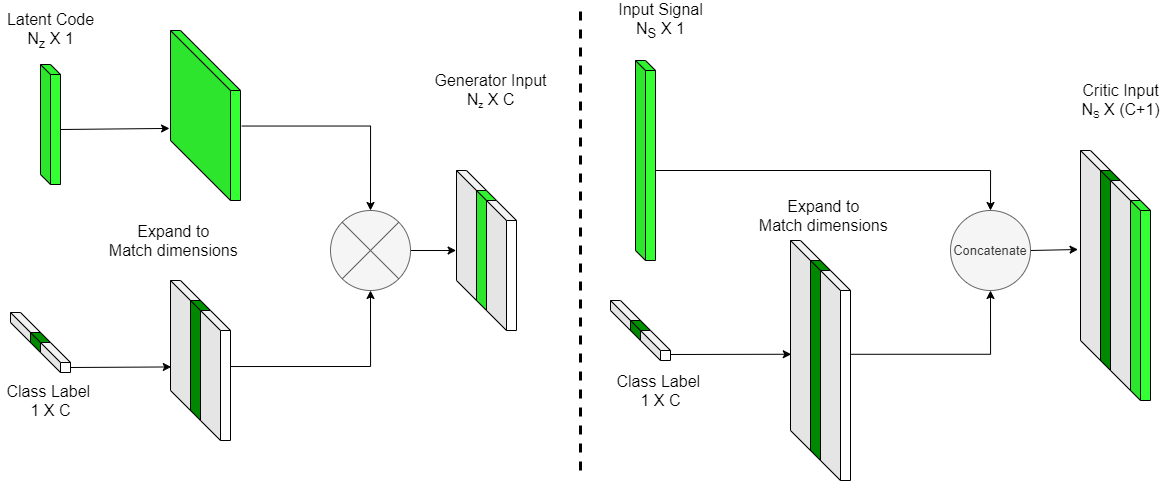}
    \caption{PowerGAN's method of conditioning the generator and critic. On the generator side (left), the input latent code and the one-hot class label are both extended and then multiplied. Effectively, this is equivalent to placing a copy of the latent code in the corresponding column matrix which is zero everywhere else. On the critic side (right), we perform a similar extension of the class labels, but then simply concatenate the resulting tensor to the input signal.}
    \label{fig:conditioning} 
    \vspace{-0.3cm}
\end{figure*}

A major novelty in PowerGAN is the introduction of conditioning, both for the generator and the critic, on the desired appliance label. Following the concepts presented in~\cite{cgan}, we choose to condition our GAN on the input labels by including the class label as an input to both the critic and the generator. On the generator side this is done by replacing the latent code input with $Z \in \mathbb{R}^{N_z \times C} = [\boldsymbol{z}^T_0, \boldsymbol{z}^T_1,...,\boldsymbol{z}^T_C]$ such that:
\begin{equation}
    \boldsymbol{z}^T_i = \begin{cases}
    \boldsymbol{z}^T & i=l \\
    \boldsymbol{0}^T & \text{otherwise}
    \end{cases}
\end{equation}
where $N_z$ is the latent space dimension, $\boldsymbol{z} \in \mathbb{R}^{N_z}$ is the latent code, $C$ is the number of different labels in the dataset, and $l$ is the current label. In practice, this is performed by extending both the latent code and the one-hot labels to $\mathbb{R}^{N_z \times C}$ and multiplying the resulting tensors. To accommodate for the added capacity required by the conditional generator, we increase the amount of features in the input stage by a factor of $C$ compared with the rest of the network. On the critic side, we simply extend the one-hot labels to $\mathbb{R}^{N_s \times C}$, where $N_s$ is the current signal length, and concatenate the resulting tensor to the input signal, as illustrated in Fig.~\ref{fig:conditioning}.

In PowerGAN, we also adopt many of the smaller, nuanced, practices proposed in~\cite{ProgGAN,eeggan}. As suggested in~\cite{ProgGAN}, 
to alleviate growing magnitude issues, we strictly normalize each time-step in each feature map to have an average magnitude of $1$. To improve convergence during training, we employ on-line weight scaling (instead of careful weight initialization). To increase the variation of generated signals, we use a simplified version of minibatch discrimination, as proposed in~\cite{ProgGAN} and modified in~\cite{eeggan}, wherein the standard deviation is used as an additional feature for the final layer of the critic. The minibatch standard deviation is calculated first at each feature, at each time-step, and then averaged across both features and time to give one single value for the entire batch. 

Furthermore, we use the weighted one-sided variation of the gradient penalty, as proposed in~\cite{eeggan}, 
and modify it to accommodate the conditional critic and generator. The gradient penalty's importance, as noted in~\cite{eeggan}, depends on the current value of the Wasserstein distance $D_W = \mathbb{E}_{x_g}[D_\alpha(x_g,l)] - \mathbb{E}_{x_r}[D_\alpha(x_r,l)]$. When $D_W$ is large, it is important to ensure that the cause isn't the loss of the 1-Lipschitz constraint. However, when the $D_W$ is 
low, it is worthwhile to focus on optimizing it directly, and assign a lower weight to the gradient penalty. In practice, this is achieved by giving an adaptive weight to the gradient penalty equal to the current $D_W$. It is important to note that this weight is treated as a constant for gradient purposes, to avoid undesirable gradients. The gradient penalty itself is one-sided, meaning it allows for the critic to have a smaller than 1-Lipschitiz constraint, as was considered but ultimately not chosen in~\cite{wgan-gp}. In this form the gradient penalty becomes: 
\begingroup\makeatletter\def\f@size{9}\check@mathfonts
\begin{equation} 
    \mathcal{L}_{GP} = \lambda\cdot\max(0,D_W)\cdot \mathbb{E}_{\tilde{x} \sim P_{\tilde{x}}}\bigr[\max\bigl(0,\lVert\nabla_{\tilde{x}}   D\left(\tilde{x},l\right) \rVert _2 -1\bigr)^2 \bigl] 
\end{equation}
\endgroup
where $D_W$ is the current critic estimate of the Wasserstein distance, $D$ is the critic, 
 and $\tilde{x}$ is a randomly weighted mixture of pairs of real and generated samples, each with the same label $l$. Remember that $D_W$ here is treated as a constant for back-propagation purposes.

Finally, we use a small loss component to center critic output values around zero, also introduced in EEG-GAN~\cite{eeggan}:
\begin{equation}
    \label{eq:cen}
    \mathcal{L}_{C} = \epsilon \cdot\bigl(\mathbb{E}_{x_r}[D(x_r)] + \mathbb{E}_{x_g}[D(x_g)] \bigr)
\end{equation} 
where $\epsilon \ll 1$, and $x_r, x_g$ are real and generated samples, respectively. This loss helps with numerical stability as well as interpretation of the loss value during training. Combining all of the above, the final loss functions of the critic ($\mathcal{L}_D$) and the generator ($\mathcal{L}_G$) in PowerGAN are:
\begin{gather}
    \mathcal{L}_D = \mathbb{E}_{x_g}[D_\alpha(x_g,l)] - \mathbb{E}_{x_r}[D_\alpha(x_r,l)] + \mathcal{L}_{GP} + \mathcal{L}_C \\
    \mathcal{L}_G = -\mathbb{E}_{x_g}[D_\alpha(x_g,l)]
\end{gather}


Another important difference between PowerGAN and~\cite{eeggan} is in the method of resampling the signals. In~\cite{eeggan}, after comparing various methods, the authors use strided convolutions for downsampling in the critic, average pooling for downsampling the input data, and either linear or cubic interpolation for upsampling in the generator. We find that given the quick switching nature of appliance power traces, it is important to allow for high frequency changes in the signal, even at the price of some aliasing. For this reason we downsample the input signals using maxpooling, and perform the upsampling steps in the generator with nearest-neighbour interpolation.

\vspace{-0.3cm}
\subsection{Training}
\label{subsec:train}

PowerGAN  was trained using the REFIT~\cite{REFIT} dataset. 
REFIT consists of power consumption data from 20 residential homes, at the aggregate and appliance level, sampled at 1/8~Hz. The REFIT dataset was prepared by following the prescription of some recent work to ensure consistent sampling~\cite{murray_icassp}.
Because not all of the 20 houses contain the same appliances, we chose appliances that were available in multiple houses. We also wanted to ensure these appliances exemplified each of the four appliance types as defined by~\cite{hart}, and then expanded by~\cite{novel}: ON-OFF, Multi-state, Variable Load, and Always-ON (or periodic). Of the appliances available in REFIT, five that satisfied the above considerations were used: refrigerators (along with freezers, and hybrid fridge-freezers), washing machines, tumble dryers, dishwashers, and microwaves.
Each instance of these five appliances were arranged into approximately five hour windows,
centered around the available activations. We located these activations by first-order differences in power that were larger than 50 Watts. 


Windows were then filtered according to two conditions: First, the energy contained in the window should be appreciably larger than the ``steady-state'' contribution to the energy (taken here to be the sum of the window mean and half the window standard deviation). In other words, after ignoring the samples less than this value, the remaining energy contained in the window should be above some threshold, set in our work to be $33.33$ Watt-hours. This condition ensures that low-energy windows, where the activation was falsely detected due to sensor noise, are excluded. This condition also filters out windows that may contain significant energy, but have little useful structural information - mainly windows composed of a constant level of power.  

Secondly, we calculate the Hoyer sparsity metric~\cite{hoyer}, $S$, for $\boldsymbol{\delta}(w_i)$ - a vector of length $n$ containing the discrete first-order differences in each window $w_i$:
\begin{equation}
\label{eq:edge-spar}
S_{\boldsymbol{\delta}(w_i)} = \frac{\sqrt{n} - \frac{\lVert \boldsymbol{\delta}(w_i)  \rVert_1}{\lVert \boldsymbol{\delta}(w_i) \rVert_2}}{\sqrt{n}-1}
\end{equation}
where $\lVert \boldsymbol{\delta}(w_i)  \rVert_1$ and $\lVert \boldsymbol{\delta}(w_i)  \rVert_2$ are the $\ell_1$ and $\ell_2$-norms of $\boldsymbol{\delta}(w_i)$, respectively. At its extremes, the Hoyer sparsity metric is zero when every sample in $\boldsymbol{\delta}(w_i)$ is the same (meaning the $\ell_1$-norm is larger than the $\ell_2$-norm by a factor of $\sqrt{n}$), and unity when there is only one non-zero sample in $\boldsymbol{\delta}(w_i)$ (i.e., highly sparse). 
By requiring the sparsity metric to be larger than $0.5$, we ensure that windows are not overly noisy, further maximizing the structural information contained in them. The remaining windowed dataset was then balanced and the windows belonging to each appliance were normalized. 

Finally, before every epoch, windows were shifted randomly in time 
to avoid biasing the network towards specific activation locations within each window. The shifted windows were then downsampled to match the resolution of the current training stage. We utilized the Adam~\cite{adam} optimizer for training PowerGAN, setting $lr = 0.001$ and $\beta = (0,0.99)$ We trained each stage of PowerGAN for 2000 epochs, out of which the first 1000 included fading with linearly changing weights. See Algorithm~\ref{alg:powergan} for full details. 

\vspace{-0.3cm}
\section{Experiments}
\label{sec:eval}

We present both a qualitative analysis of the PowerGAN-generated power traces as well as their quantitative evaluation, based on adaptations of commonly used GAN evaluation methods to 1-D power traces. We compare quantitative metrics with two other appliance power trace synthesizers: SynD~\cite{SynD}, and ANTgen \cite{ANTgen}, which is a more up-to-date version of  AMBAL. 
SmartSim~\cite{SmartSIM} is not included in the comparison because 
the published sample data is of insufficient size for accurate comparison with other methods in these experiments. 

When generating signals 
using PowerGAN, we found it beneficial to add two simple post-processing steps: we ensure that at any given time-step the generated power is larger than zero; and we discard any generated signals that do not meet the energy threshold designated for the training data (and replace them with new generated samples). 
\setlength{\textfloatsep}{0pt}
\begin{algorithm}[ht]
\caption{PowerGAN Training Procedure}
\label{alg:powergan}
\begin{algorithmic}[1]
\Require Real samples with corresponding labels $(x_R,l)\in X_R$; Conditional Generator $G(\boldsymbol{z},l)$; Conditional Critic $D(\boldsymbol{x},l)$; optimizers for $G,D$.
\Ensure $N_b$: Number of blocks for $G,D$; $EP_b$: number of training epochs per block; $EP_{f}:$ number of fading epochs; $R$: ratio of critic to generator training iterations.
\For{$n = 1,2, \ldots, N_b$}
\State Add Block to $G,D$
\For{$ep = 1,2, \ldots, EP_b$}
\State Set $\alpha = \min(1,ep/EP_{f})$
\State Set $G_\alpha, D_\alpha$ according to Fig.~\ref{fig:fading}
\State Randomize appliance starting points
\LeftComment and downsample $X_R$ by $2^{N_b-n}$
\State Select a minibatch of real samples and labels: $\boldsymbol{x}_R,\boldsymbol{l}$
\State Generate a mini-batch of samples using 
\LeftComment labels: $\boldsymbol{x}_G = G_\alpha \bigl(\boldsymbol{z}\backsim \boldsymbol{N}(0,\mathbb{I}) , \boldsymbol{l}\bigr)$
\State $\mathcal{L}_D = \mathbb{E}_{x_g}[D_\alpha(x_g,l)] - \mathbb{E}_{x_r}[D_\alpha(x_r,l)] + \mathcal{L}_{GP} + \mathcal{L}_C$
\State Take optimizer step for D
\If {$ep==0\mod{R}$}
\State generate a mini-batch of samples using 
\LeftComment labels: $\boldsymbol{x}_G = G_\alpha \bigl(\boldsymbol{z}\backsim \boldsymbol{N}(0,\mathbb{I}) , \boldsymbol{l}\bigr)$
\State $\mathcal{L}_G = -\mathbb{E}_{x_g}[D_\alpha(x_g,l)]$
\State Take optimizer step for G
\EndIf
\EndFor
\EndFor
\end{algorithmic}
All expected value operations are approximated using the sample mean of the minibatch.
\end{algorithm}

\begin{figure*}[htbp]
  \centering
    \includegraphics[width = \textwidth]{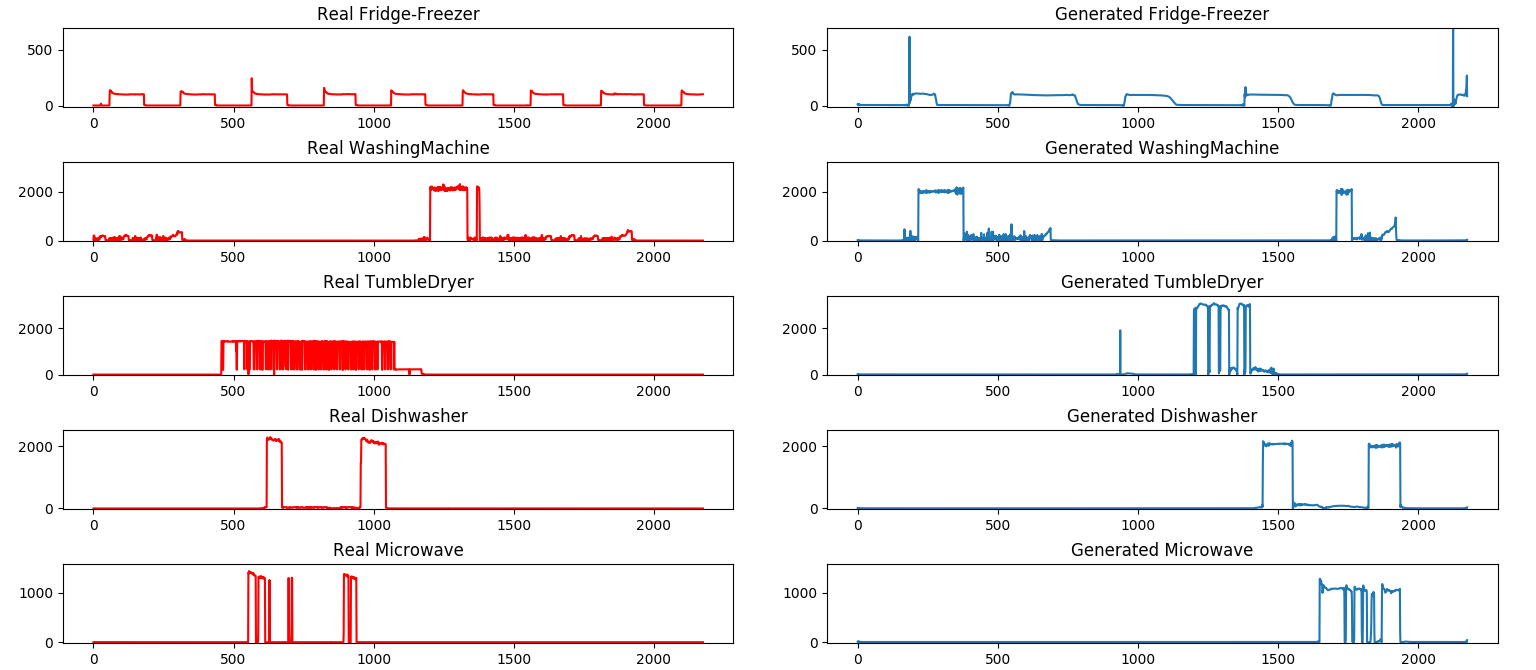}
    \caption{Examples of appliance power traces generated by PowerGAN, alongside their real counterparts taken from REFIT. We can see here that the generated signals follow the real data closely, yet without direct copying, in important attributes such as power levels, overshoot, quick switching, and more.}
    \label{fig:all_apps} 
    \vspace{-0.3cm}
\end{figure*}

\vspace{-0.3cm}
\subsection{Quantitative Evaluation}
\label{subsec:quantity}

Tasks such as segmentation, classification, regression, or disaggregation, are relatively easy to evaluate because they have a well-defined goal. While there are several different approaches to evaluating NILM~\cite{NILMeval}, all methods utilize a well-defined ground truth, such as appliance power consumption or state. Unfortunately, no such ground truth exists when attempting to evaluate randomly generated signals. In fact, the attempt to assign a numerical value to measure the quality of a GAN framework is in itself a significant and challenging research problem~\cite{GAN-eval}. To evaluate PowerGAN, we choose three commonly used GAN evaluation metrics, and adapt them to be applicable for power trace data. 

Inception score (IS)~\cite{improved} uses a pre-trained DNN-based classifier named Inception \cite{inception}, to evaluate the quality of generated signals. To calculate IS, a batch of generated samples are classified using the pre-trained model. The output of this classifier can be seen as the probability that a sample belongs to each target class. A good generator is realistic, meaning we expect low entropy for the output of the classifier. Simultaneously, a good generator is also diverse, meaning we expect high entropy when averaging out all classifier outputs. To include both requirements in one numerical measure,~\cite{improved} defines the Inception score as $IS = \exp\Bigl(\mathbb{E}\bigl[D_{KL}\bigl(p\left(y\vert \boldsymbol{x}\right) \Vert~  p\left(y\right)\bigr)\bigr]\Bigr)$, where $D_{KL}$ is the KL divergence. 

Because the IS is not an objective metric, it is common to compare the generator's score with the score obtained from real data. Because no such classifier is commonly used for power trace signals, we train our own model, using a one dimensional ResNet~\cite{resnet} architecture. To avoid biasing the model towards PowerGAN we also include training data from ECO~\cite{ECO} and Tracebase~\cite{Tracebase}, as they were the foundation used for the ANTgen power traces. The real power traces, used as foundation for SynD, were not published, so they could not be included in classifier training. We then evaluate the IS in batches and present the mean and standard deviation for each generator, as well as the real data.

While IS has shown good correlation with human classification of real versus generated samples, it is not without its flaws. It is highly sensitive to noise and to scale, as well as mode collapse. For example, if a model can generate exactly one, highly realistic, sample for every class, it will achieve near perfect IS, without actually being a diverse generator. To avoid some of these pitfalls,~\cite{frechet} introduced the Frechet Inception Distance (FID). The FID uses the same classifier as IS, but instead of measuring probabilities directly at the output, it evaluates the distributions of features in the final embedding layer of the classifier. FID measures the Wasserstein 2-distance between the distribution of real and generated signal features, under a Gaussian assumption (which allows a closed-form solution). The FID is significantly less sensitive to mode collapse and noise, yet still struggles with models that directly copy large portions of the training set. Because FID is a proper distance, its value can serve as a more objective metric. We evaluate FID using the full set used for training our ResNet classifier, and generate an equivalent amount of data from each synthesizer. 

A similar approach to FID, the sliced Wasserstein distance (SWD)~\cite{ProgGAN} attempts to evaluate the difference between the distributions of real and generated signals directly. SWD uses 1-D projections to estimate the Wasserstein distance between two distributions, taking advantage of the closed form solution for the distance of such projections. In practice, the SWD is itself approximated using a finite set of random projections. It is common to evaluate SWD on some feature space, to make it more robust. For our work, we compare two possible feature sets: the classifier features used for FID, and a Laplacian ``triangle" (a 1-D adaption of a Laplacian pyramid) using a 15-sample Gaussian kernel. Similarly to FID, we evaluate the SWD on the entire training set, and we use 10 iterations of 1000 random projections each, calculating the mean and standard deviation along the iterations. Table~\ref{tbl:NumericEval} summarizes the results for all the metrics described above.

\begin{table}[htbp]
\caption{Synthesized Appliance Performance Evaluation}
\label{tbl:NumericEval}
\vspace{-0.5cm}
\begin{center}
\begin{tabular}{|c|c|c|c|c|}
\hline
\textbf{Generator} & $\boldsymbol{IS}$ &  $\boldsymbol{FID}$ & $\boldsymbol{SWD}_{Lap}^*$ & $\boldsymbol{SWD}_{Cl}$ \\
\hline
Dataset & $3.77 \pm .15$ & 0 & 0 & 0\\
\hline
ANTgen & $ 3.73  \pm .11$ & $69.63$ & $45  \pm .029$ & $0.31  \pm .017$\\
SynD & $3.18  \pm .10$ & $76.09$ & $ 22  \pm .011$ & $0.33  \pm .015$\\
PowerGAN & $\boldsymbol{3.81  \pm .13}$ & $\boldsymbol{43.30}$ & $\boldsymbol{18  \pm .088}$ & $\boldsymbol{0.25  \pm .011}$\\
\hline
\end{tabular}
\end{center}
$^{*} SWD_{Lap}$ values were calculated using Laplacian ``triangle" features were scaled by $10^{-3}$. $SWD_{Cl}$ values were calculated using the last layer of classifier features, similarly to the Frechet Inception distance.
\vspace{-0.3cm}
\end{table}

Several things stand out when reviewing the quantitative results. First, we notice PowerGAN receives the highest Inception score, outscoring both SynD and ANTgen in a statistically significant manner (t-test  $p\leq 1e^{-5}$). PowerGAN even slightly outscores  the real data, although not in a statistically significant manner (t-test $p=0.38$). We believe this is caused by the existence of some inevitably mislabeled data in REFIT. When collecting sub-meter data for NILM applications, the wiring of certain houses makes it difficult to avoid having more than one appliance on each sub-meter. This means that often a sub-meter designated as one appliance (such as fridge or dishwasher) will contain measurements from a smaller, or less commonly used appliance (such as a kettle or battery charger). The presence of such activations may lead to a lower Inception score in the real data, but effects PowerGAN to a lesser extent. 

Secondly, we notice that the diversity of PowerGAN-generated signals is noticeable when reviewing the more advanced metrics. In both variations of the $SWD$ as well as FID, PowerGAN outperforms the other two synthesizers in a statistically significant manner (t-test $p \leq 9e^{-4}$). 
We believe that the combination of these scores shows that PowerGAN is capable of generating samples that are comparable, in terms of realism, with copying or hand-modeling real data directly (as done by SynD and ANTgen), while at the same time creating diverse and truly novel appliance power signatures.

\vspace{-0.3cm}
\subsection{Qualitative Analysis}
\label{subsec:quality}

\begin{figure*}[htbp]
  \centering
    \includegraphics[width = 0.85\textwidth]{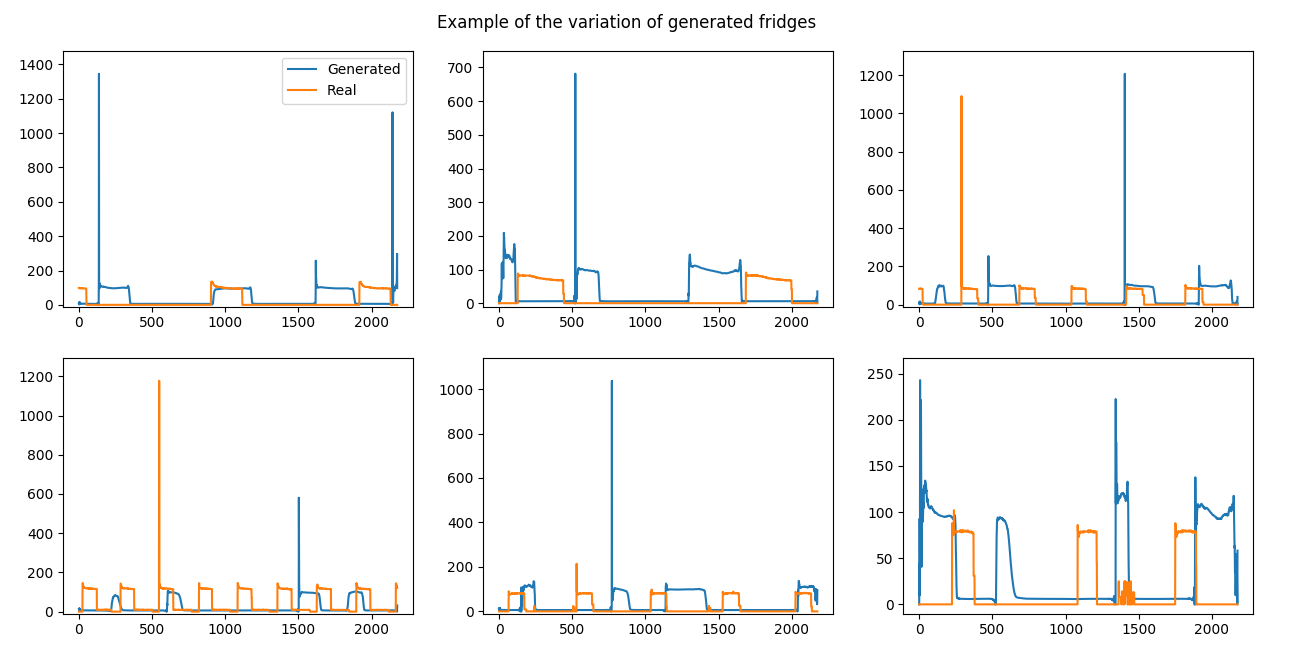}
    \caption{Examples of generated and real fridges. There is diversity in the generated fridges in terms of frequency, duty cycle, overshoot size, and more. PowerGAN generates some artifacts such as an overshoot at the end of an activation, as well as some power variations within a given activation.}
    \vspace{-0.3cm}
    \label{fig:multigen} 
\end{figure*}

When evaluating our generated signals, we focus on the traces' realism as well as their variety and novelty. We find that PowerGAN is able to generate highly realistic-looking appliance traces while avoiding directly copying existing appliances from REFIT. In addition, we notice that the generator's diversity exists both between classes and within each class. 

Fig.~\ref{fig:all_apps} shows an example of generated signals from each of the five trained appliances, along with similar real power traces. We can see that the generated signals present highly comparable behaviours and contain all of the major features of each appliance class. Some important attributes in the generated signals are shown below, by class:


\vspace{-0.15cm}
\begin{itemize}
    \item \textbf{Fridges} - generated fridge traces maintain the periodic nature of real refrigerators. We see small variation in both frequency and duty cycles of the activations, with minor differences within an activation and larger differences between different samples. In addition, generated fridges maintain the initial spike in power consumption.
    \item \textbf{Washing Machines} - generated washing machine traces manage to convey the complicated state transitions of the various washing cycle states. We see quick fluctuations in power consumption, typical of the machine's internal heating unit switching on and off. Additionally, the generator is able to generate the variable load which occurs during the washing machine's spin cycle.
    \item \textbf{Tumble Dryers} - generated tumble dryer traces are able to maintain the characteristic drop in power consumption that occurs periodically when the dryer changes direction. Furthermore, PowerGAN is able to capture the usage characteristics of a dryer, occasionally including more than one activation in a 5-hour window.
    \item \textbf{Dishwashers} - generated dishwasher traces manage to maintain the multi-state properties of the original dishwashers, without incurring significant amount of switching noise or any major artifacts.
    \item \textbf{Microwaves} - generated microwave traces portray the low duty cycle of real microwaves, which are generally only used occasionally for periods of a few minutes at most. In addition, PowerGAN is able to generate traces that include quick switching of the microwave oven, which can occur during more advanced microwave modes such as a defrost program.
\end{itemize}

While 
PowerGAN generates realistic data for the most part, some issues still exist. The generated signals occasionally contain artifacts that are rare in real signals, such as an overshoot before deactivation, power fluctuations within a given state, or unlikely activation duration. When analyzing these artifacts, we note that examples of such behaviour exist in the real data, albeit rarely. We believe that these behaviours appear in PowerGAN because in the training procedure, such artifacts become central in identifying appliances, leading to them carrying significant gradients to the generator. 

In order to demonstrate the diversity of the power traces generated by PowerGAN, we present six examples of generated and real fridge signals in Fig.~\ref{fig:multigen}.  
We note that like the real fridge power traces, the generated signals vary in several important features: power level, activation frequency, duty cycle, and overshoot size. In addition, the generated signals demonstrate some variations in each of the above parameters within an activation window, similarly to real fridges.

\section{Conclusions}
\label{sec:Summary}

After identifying the need for synthetic data generation for NILM, we presented here the first GAN-based synthesizer for appliance power traces. Our model, named PowerGAN, is trained in a progressive manner, and uses a unique conditioning methodology to generate multiple appliance classes using one generator. We have also implemented some groundwork for evaluating power trace generators which, as expected, requires more than one metric in order to evaluate the various requirements from synthesizers. Using these metrics, along with visual inspection of the generated samples, we have shown that PowerGAN is able to produce diverse, realistic power appliance signatures, without directly copying or hand-modeling the training data.

While the results presented in this paper are based on training on the REFIT dataset, the presented framework can be used for training on any desired dataset, and at any sampling frequency. We believe that these properties may help researchers in using PowerGAN as an augmentation tool for training supervised NILM solutions. The PowerGAN generator can be used to randomly replace certain activation windows in the real data with synthesized ones, with the hope of improving 
out-of-distribution performance. In order to do this, one can modify the training procedure of PowerGAN slightly to include the desired activation window sizes, as well as remove the random time shifting during training, if a well localized activation is preferred for disaggregation.

\bibliographystyle{IEEEtran}
\bibliography{bare_jrnl}




\ifCLASSOPTIONcaptionsoff
  \newpage
\fi

\end{document}